\documentclass[twocolumn]{aastex62}

\usepackage{textcomp}
\usepackage{palatino}

\graphicspath{{./}{figures/}}

\shortauthors{Markkanen et al.}
\begin{document}

%\title{Interpretation of phase functions of the comet 67P/Churyumov-Gerasimenko measured by the OSIRIS instrument }

\title{Interpretation of the phase functions measured by the OSIRIS instrument for Comet 67P/Churyumov-Gerasimenko}

\correspondingauthor{Johannes Markkanen}
\email{markkanen@mps.mpg.de}

\author[0000-0001-5989-3630]{Johannes Markkanen}
\affiliation{Max Planck Institute for Solar System Research, Justus-von-Liebig-Weg 3, 37077 G\"{o}ttingen, Germany}

\author{Jessica Agarwal}
\affiliation{Max Planck Institute for Solar System Research, Justus-von-Liebig-Weg 3, 37077 G\"{o}ttingen, Germany}
%\collaboration{(AAS Journals Data Scientists collaboration)}

\author[0000-0003-3815-0293]{Timo V\"{a}is\"{a}nen}
\affiliation{Deparment of Physics, P.O. Box 64,
FI-00014 University of Helsinki, Finland}

\author[0000-0001-7403-1721]{Antti Penttil\"{a}}
\affiliation{Deparment of Physics, P.O. Box 64,
FI-00014 University of Helsinki, Finland}

\author[0000-0001-8058-2642]{Karri Muinonen}
\affiliation{Deparment of Physics, P.O. Box 64,
FI-00014 University of Helsinki, Finland}

\affiliation{Finnish Geospatial Research Institute FGI, Geodeetinrinne 2, FI-02430 Masala, Finland}

%\author{Homer J. Simpson}
%\affiliation{Springfield Observatory, ??,  USA}
%\affiliation{AAS Journals Associate Editor-in-Chief}
\nocollaboration



\begin{abstract}

We show that the scattering phase functions of the coma and the nucleus of the comet 67P/Churyumov-Gerasimenko measured by the Rosetta/OSIRIS instrument can be reproduced by a particle model involving clustered densely packed submicrometer-sized grains composed of organic material and larger micrometer-sized silicate grains. The simulated and measured coma phase functions suggest that near the nucleus scattering is dominated by large particles, and the size distribution of dust particles varies with time and/or local coma environment. Further, we show that the measured nucleus phase function is consistent with the coma phase function by modelling a nucleus-sized object consisting of the same particles that explain the coma phase functions. 

\end{abstract}



\section{Introduction} 

The European Space Agency's (ESA) Rosetta mission to the comet 67P/Churyumov-Gerasimenko has provided a unique opportunity to study a cometary dust environment by multiple instruments. Understanding the dust environment is crucial for the interpretation of the remote light-scattering observations of comets in general. The light-scattering features such as the intensity phase function and the degree of linear/circular polarization of the coma and the nucleus depend on the physical properties of dust, i.e., the shape and size of particles, porosity, and chemical composition of cometary material. In the following, we refer to the constituting monomers as {\it grains}, and to aggregates of these grains as {\it particles}.

The OSIRIS (Optical, Spectroscopic, and Infrared Remote Imaging System) camera on board the Rosetta spacecraft allowed for absolute magnitude imaging, and the images have been used to construct the intensity phase functions of the coma \citep{Bertini2017} and the nucleus \citep{Fornasier2015, Feller2016, Nafiseh2017}. The MIDAS (Micro-Imaging Dust Analysis System) and COSIMA (COmetary Secondary Ion Mass Analyser) instruments have provided clues for morphology of dust. The atomic force microscope of MIDAS shows evidence that the dust particles are agglomerated submicrometer-sized irregular grains with the equivalent radius $r \sim 100 - 1000$\,nm \citep{Bentley2016, Mannel2016}. Albeit, the resolution limits the detection of even smaller sub units hence their existence cannot be excluded. COSIMA has captured and imaged a huge number of large particles ranging from 10\,\textmu m  up to 1\,mm suggesting that large particles dominate the scattering features of the coma near the nucleus as MIDAS has only detected few particles smaller than 10 \textmu m. The COSIMA particles may be agglomerates of smaller units, possibly composed of similar particles detected by MIDAS \citep{Hilchenbach2016,Langevin2016}. Further, COSIMA has analyzed the elemental and isotopic composition of the surface of the dust particles by the secondary ion mass spectrometer. The analysis indicates that organic material and silicate minerals are the major components of dust mixed together in a scale smaller than the resolution of the instrument ($35 \times 50$\,\textmu m$^2$) \citep{Bardyn2017}.

Regardless of the knowledge on the morphology and composition of the dust particles obtained by the various in situ measurements, the light-scattering characteristics of the coma and the nucleus remain unexplained. Especially, the deep minimum in the intensity phase function around 100$^{\circ}$ \citep{Bertini2017} cannot be reproduced with the commonly used cometary dust particle models such as the fractal aggregates of spherical particles \citep{Kimura2006} or agglomerated debris particles \citep{Zubko2009}. \cite{Moreno2018} showed that aligned spheroidal particles much larger than the wavelength reproduce the coma phase function. However, the model fails to explain the ground based polarimetric observations, and the mechanism yielding such alignment has not been demonstrated in practice.  

In this paper, we show that the model based on randomly oriented large dust particles consisting of submicrometer-sized organic grains and micrometer-sized silicate grains explains the observed coma phase functions. The model also suggests that the variations in the phase functions with time are due to the different particle size distribution. In addition, the model predictions are consistent with the ground-based polarimetric observations. Finally, we show that the nucleus phase function can be reproduced by a surface consisting of densely packed particles with the same scattering properties as the coma particles.

\section{Numerical method} 
To model light scattering by large particles consisting of wavelength-sized grains is impossible with the standard numerical light-scattering methods such as the integral-equation, finite-element or finite-difference methods. We have recently introduced the radiative transfer with reciprocal transactions method (R$^2$T$^2$) \citep{Muinonen2018,Markkanen2018} which extends the applicability of the radiative transfer to the dense discrete random medium of low absorbing particles. Here we include the coherent component to the method using the mean field correction which allows us to treat highly absorbing particles.  

The mean field correction takes the refractions and reflections of the mean field (coherent field) into account using Snel's law. The correction is crucial for a medium consisting of highly absorbing grains such as cometary dust as the mean free path is short, and consequently, the surface of the particle has a significant effect on scattering. The drawback of the mean field correction is that taking the coherent backscattering effect arising from the volume element interactions into account is not trivial and it is excluded in this paper. We note, however, that the coherent backscattering contributes only at the small phase angles and is typically assumed to be rather weak for dark material.

\begin{figure}[htb]
\begin{center}
\includegraphics[width=0.5\textwidth]{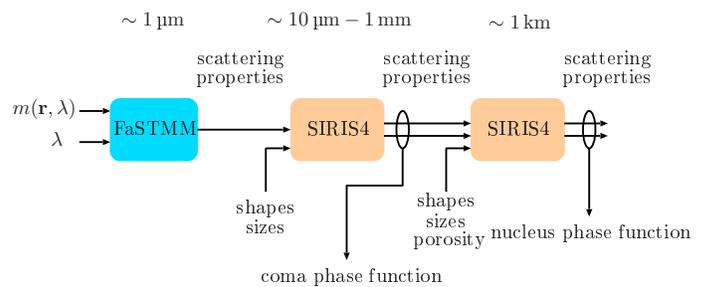}
\caption{Illustration of the computational chain:  First, the FaSTMM solver computes the scattering properties of the volume elements containing large number of small grains. These are inserted into the SIRIS4 solver which outputs scattering properties of dust particles. The scattering properties of the dust particles are used as input to the SIRIS4 which gives scattering properties of the nucleus as output.}
\end{center}
\label{fig0}
\end{figure}

Computations proceed as follows: First, we compute and store a large number of scattering properties of volume elements. A volume element contains a large number of randomly oriented and positioned grains characterized by the position $\mathbf r$ and wavelength $\lambda$ dependent refractive index function $m(\mathbf r, \lambda)$. Thus, they capture the statistics of the random medium and treat the near-zone interactions rigorously. In the computations, we apply a numerically exact fast superposition $T$-matrix method (FaSTMM) \citep{Markkanen2017}. Second, we calculate the incoherent and coherent scattering characteristics of the volume elements. From the coherent scattering properties we extract the real part of the effective refractive index for the coherent field by matching the coherent scattering cross section to that of a homogeneous sphere with the effective refractive index. The ensemble averaged mean free path, phase function, and albedo are calculated from the incoherent scattering properties of the volume elements. The imaginary part of the effective refractive index for the coherent field is linked to the extinction coefficient of the incoherent field as it describes the rate of energy transfered from the coherent to incoherent component \citep{Zurk1996}. 
Finally, we apply the combined geometric optics and radiative solver (SIRIS4) \citep{Lindqvist2018} to calculate scattering by large dust particles using the effective refractive index and the incoherent scattering properties of the volume elements as input \citep{Martikainen2018}. This procedure allows us to analyze scattering by large particles consisting of wavelength-sized grains. Further, using SIRIS4 in a hierarchical manner, i.e., using output as input in the second round, we can compute scattering by the nucleus consisting of large dust particles. The entire computational chain is illustrated in Figure \ref{fig0}.

\section{Coma phase function}

To model the coma phase function, we assume that the coma is optically thin and the dust particles are in each other's far zone. Consequently, multiple scattering effects between the particles can be neglected and the scattering properties of the coma can be computed by averaging over an ensemble of particles. It is important to note that, since we do not know the optical thickness of the coma, we cannot compare the absolute amplitude of the modeled and measured phase functions. We normalize the modeled phase functions to the geometric albedo at zero phase angle in order to preserve the absolute scale. The measured phase functions, in turn, are normalized to the averaged modelled values at $\alpha = 20^{\circ}$, that is the lowest phase angle available for all measurements. 

We have computed scattering by particles consisting of organic, silicate, and ice grains with varying grain sizes ($r = 50$ -- 1000\,nm) and packing densities ($v = 0.1$--0.4). Our simulations show that a single grain population cannot reproduce all the measured phase functions, typical polarization curves of comets, and the estimated geometric albedo of dust particles. Specifically, obtaining the deep minimum in the intensity phase function around $ \alpha = 100^{\circ}$ and the strong opposition effect is challenging with a one-component particle model. Thus, we introduce two different grain populations occupying dust particles: The first population contains submicrometer-sized ($r = 65$ -- 125\,nm) highly absorbing spherical grains with the refractive index $m_1=2.0 + \rm{i}0.2$ at $\lambda = 0.649$\,nm. The second population includes larger ($r = 0.6$ -- 1.3\,\textmu m) weakly absorbing spherical grains with $m_2=1.6 + \rm{i}0.0001$ at $\lambda = 0.649$\,nm. We assume the differential power-law index of -3 for both populations. 

The highly-absorbing grains can be associated with organic material and the weakly-absorbing grains correspond to typical silicate mineral \citep{Li1997}. The volumetric filling factor for the organic grains is $v_{\rm org}=0.3$ and $v_{\rm sil}=0.0375$ for the silicate grains. The volumetric ratio of the organic and mineral phases cannot be directly determined by COSIMA as the distribution of oxygen and the relative densities of the two phases are not known. Nonetheless, our value for the volumetric organic/silicate ratio is a reasonable estimation of the estimated organic/mineral mass fraction reported by COSIMA \citep{Bardyn2017}.

\begin{figure}[htb]
\begin{center}
\includegraphics[width=0.3\textwidth]{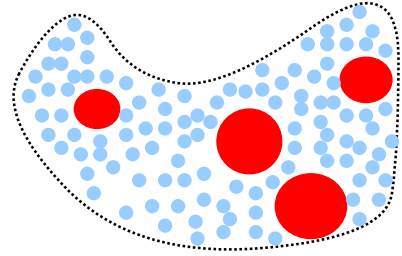}
\caption{Schematic of the particle model is shown. The blue spheres correspond to submicrometer-sized organic material and the larger red spheres represent micrometer-sized silicate grains. The shape of the entire particle is described as a Gaussian random sphere.}
\end{center}
\label{fig1}
\end{figure}

The shape of the particles is described by the Gaussian random sphere shape (GRS) \citep{Muinonen1996}. The GRS shapes are defined by the spherical-harmonics representation for the logarithmic radial distance. The statistical shape parameters are the standard deviation of the mean radius $\sigma$ and the power-law index for the covariance function $\nu$. A schematic of the particle model is presented in Figure \ref{fig1}.
 
\begin{figure}[htb]
\begin{center}
\includegraphics[width=0.5\textwidth]{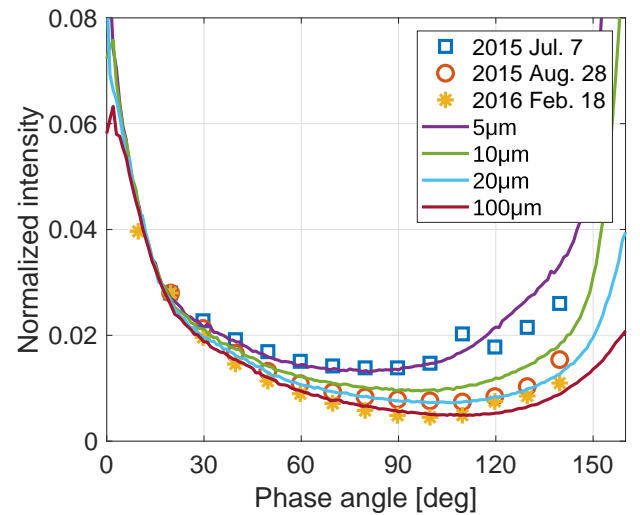}
\caption{Modelled intensity phase functions for different particle sizes are shown (solid lines). The markers show three coma phase curves for the orange filter (649.2\,nm) measured at different times \citep{Bertini2017}.}
\end{center}
\label{fig2}
\end{figure}

Figure \ref{fig2} plots the modelled intensity phase functions for different particle sizes. It is clear that, as the size of the particles increase, the minimum deepens and shifts towards higher phase angles. Thus, the variations of the measurements taken at different times can be explained by varying size distribution of dust particles in the local coma. This is consistent with \citep{Bertini2018} which shows that after perihelion the size of the dust particles in the coma decreases with increasing distance from the nucleus. We also see from the computations that the phase function in the backscattering direction is almost independent on the particle size. This is not surprising since the scattering properties near the backscattering direction are mostly dominated by the interactions of the wavelength-scale irregularities.

\begin{figure}[htb]
\begin{center}
\includegraphics[width=0.5\textwidth]{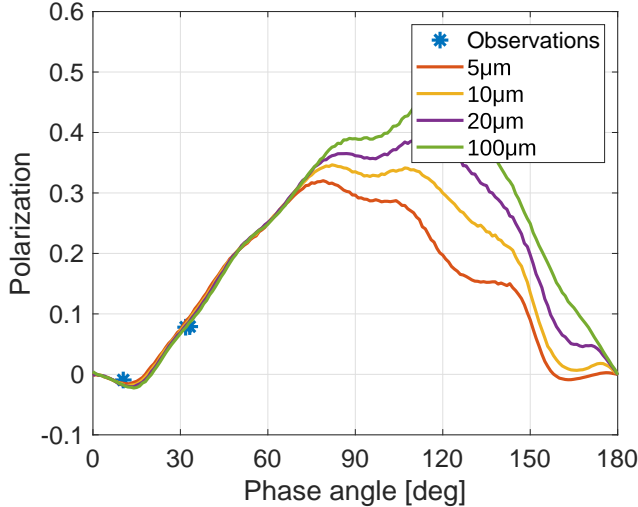}
\caption{Modelled polarization phase functions for different particle sizes, and the ground-based observations integrated over the photocenter with a radius of  1250\,km are plotted. }
\end{center}
\label{fig3}
\end{figure}

Polarimetric data is not available from the Rosetta mission, but we compare our model predictions to the ground-based observations \citep{Rosenbush2017} in Figure \ref{fig3}. As is evident from the polarization maps \citep{Rosenbush2017}, the coma is inhomogeneous and polarization varies with the distance from the nucleus. Thus, we compare our model to the polarimetric data integrated over a disk of radius 1250 km centered at the photocenter. The predicted polarization is consistent with the measurement, however, it is important to note that the measured polarization contains also the nucleus contribution.

\section{Nucleus phase function}

\begin{figure}[htb]
\begin{center}
\includegraphics[width=0.45\textwidth]{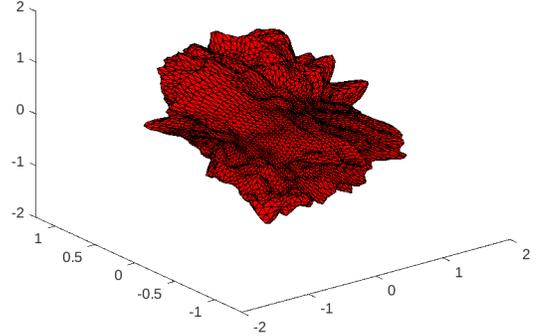}
\caption{A sample GRS shape with the standard deviation for the mean radius $\sigma = 0.2$ and the power-law index $\nu = 2.5$ for the correlation length.}
\end{center}
\label{fig4}
\end{figure}

Next, we model the scattering properties of the nucleus. As a nucleus model, we use an ensemble of 1 km-sized GRS shapes filled with smaller particles. Thus, we assume that the surface roughness follows the Gaussian statistics resulting in the corresponding geometric shadowing effect. A sample GRS particle is shown in Figure \ref{fig4}. The small particles are the same particles as we used to model the coma phase function. We apply the differential power-law size distribution of index -3 with the minimum and maximum cut off being 5\,\textmu m and 100\,\textmu m, respectively.  

\begin{figure}[htb]
\begin{center}
\includegraphics[width=0.5\textwidth]{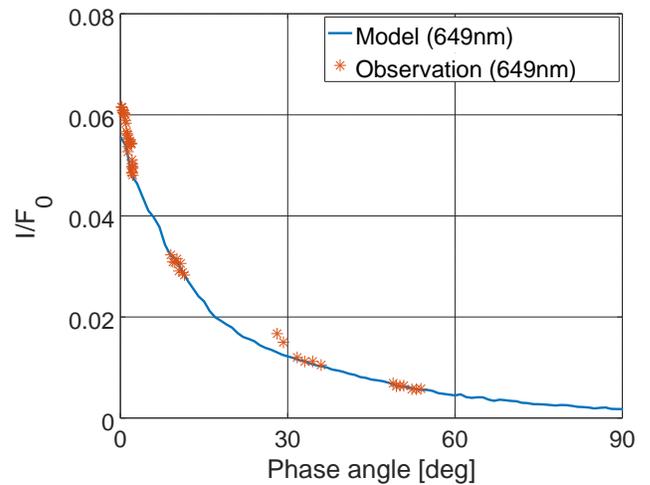}
\caption{Measured and modeled intensity phase functions for the nucleus. }
\end{center}
\label{fig5}
\end{figure}

\begin{figure}[htb]
\begin{center}
\includegraphics[width=0.5\textwidth]{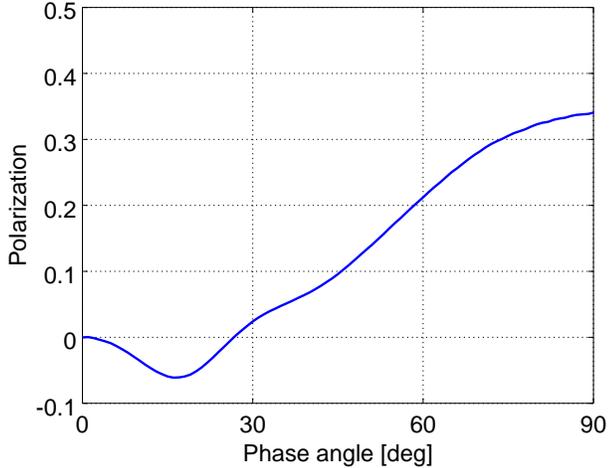}
\caption{Modelled polarization phase function of the nucleus.}
\end{center}
\label{fig6}
\end{figure}

Figure \ref{fig5} presents the modelled intensity phase function of the nucleus and the corresponding polarization is plotted in Figure \ref{fig6}. The measured data is from \citet{Nafiseh2017}. To fit the data we have assumed a packing density of 0.33 and the GRS shape parameters are $\sigma = 0.2$  and $\nu = 2.5$. The packing density has a negligible effect on the overall shape of the intensity phase function as the radiative transfer equation is independent of the packing density for extremely large objects. It mainly affects the backscattering direction due to the coherent backscattering effect. Increasing the surface roughness, i.e., decreasing $\sigma$ and increasing $\nu$, results in the steeper phase function. 
  
The overall agreement between the modelled and measured phase functions is excellent. However, the measured intensity phase function shows slightly larger value at the backscattering direction. The difference may be due to the coherent backscattering (CB) effect as our numerical scheme does not include the CB-effect arising from the volume element interactions. It only accounts for the CB effect between grains inside the volume elements. Hence there should be a small increase in the modelled intensity at the backscattering direction. Also, the measured phase function is a combined local and global  function in which the measurements at small phase angles are obtained from a local surface of the Imhotep-Ash region and may thus deviate from the integrated phase function.

\section{Wavelength dependence}

Finally, we study the behavior of the phase functions with respect to the wavelength in the visible domain. We define the spectral slope, i.e., the reddening as
\begin{equation}
S = \frac{(I/F)_2 - (I/F)_1}{[(I/F)_2 + (I/F)_1]/2}  (\lambda_2 - \lambda_1)^{-1}  
\end{equation}
where $\lambda_1 = 744$\,nm and $\lambda_2 = 481$\,nm, and $(I/F)_1$ and $(I/F)_2$ are the corresponding reflectance functions, respectively. 

We determine the refractive indices of organics at $\lambda = 481$\,nm and $\lambda = 744$\,nm by fitting the modelled reddening at $\alpha = 90^{\circ}$ to the measured coma reddening, while assuming that the refractive index for the silicate grains is constant in this wavelength range which is also a reasonable assumption for various silicates. The resulting refractive indices for organics are $m_{481 nm} = 1.9 + {\rm i}0.22$ and $m_{744 nm} = 2.05 + {\rm i}0.15$. The wavelength dependence of the refractive indices are therefore consistent with the refractive indices reported by \citet{Li1997}.

\begin{figure}[htb]
\begin{center}
\includegraphics[width=0.5\textwidth]{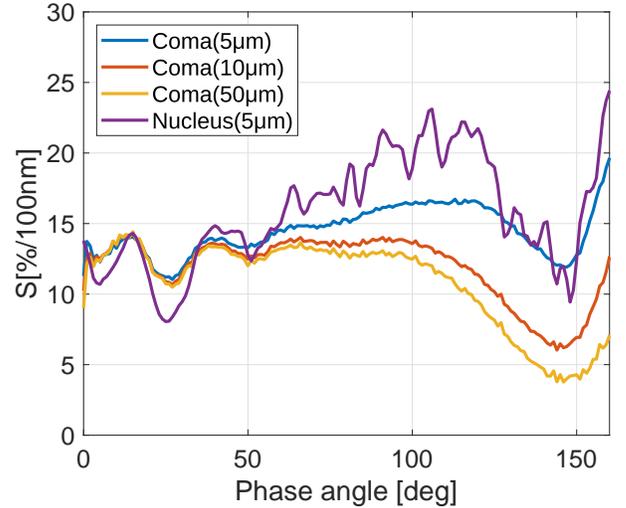}
\caption{Modelled spectral slopes for the coma and the nucleus. Values in parentheses indicate the lower cut off limits for the power-law size distribution of index -3.}
\end{center}
\label{fig7}
\end{figure}

Figure \ref{fig7} displays the reddening as a function of the phase angle for the coma particles of different size distributions and for the nucleus. We see a mild phase reddening effect for $\alpha  < 100^\circ$, when small particles dominate scattering in the coma. When larger particles dominate scattering, the phase effect vanishes. \citet{Bertini2017} has found none or little phase reddening in the coma. The nucleus shows larger phase reddening than the coma. The oscillations near the backscattering direction are caused by the Mie resonances of large spherical particles. Hence, they should be absent from the real dust measurements.

\section{Discussion} 
The coma phase functions retrieved from the images taken by the OSIRIS camera near the nucleus show large variations of the location and the deepness of the minimum of the intensity phase functions. Our simulation results suggest that a possible explanation is that the dust particle size distribution varies between different measurement series. The modelled minimum becomes deeper and shifts towards larger phase angles with increasing particle size. Also, the phase reddening becomes negligible with large particles. 

In order to obtain the deep minimum, strong opposition effect, and the realistic albedo and polarization, bright micrometer-sized mineral grains need to be present in the particles consisting mostly of organic submicrometer-sized grains (volumetric organic/silicate $\approx$ 7/1). Pure organic grains do not reproduce the strong enough opposition effect while too many silicate grains result in unrealistic polarization features and too high albedo. We note, however, that if the silicate grains are covered by organic material, the contribution of the silicate grains to the scattered light is negligible and the silicate volume fraction may be much higher. 

The size distribution of the grains has a major impact on the scattering features. Increasing the size of organic grains results in a flatter phase function in the mid phase angles ($60$--$120^{\circ}$). The silicate grains should be large enough to obtain the strong opposition effect. The power-law index and the cut off limits affect the wavelength dependence. The differential power-law index around -3 gives rise to the observed wavelength dependence, i.e., negligible phase reddening for large particles. We note, however, that the grain sizes derived in this work should not be considered as the exact but rather directional values, as the solution for the inverse problem for the Maxwell equations is not unique, and thus, the grain shape and the refractive index affect the retrieved sizes. 

The porosity and fractality of particles have a minor impact on the shape of the phase function assuming that the particles are large enough, i.e., much larger than the mean free path length. They mainly affect  scattering near the backscattering direction. Thus, to retrieve more detailed constraints on the porosity and morphology of the particles, requires accurate measurements and modelling of the scattering features near the backscattering direction.   

%{\bf (JA) Could you briefly discuss if and how a potential fractality of the grains would affect the model results?}

The ground-based observations of typical comets show a much shallower phase function minimum around $\alpha = 55^\circ$ compared to the OSIRIS coma phase functions. Such a phase function can be explained by smaller particles. This may indicate that large particles cannot escape from near the nucleus or they break apart, and therefore, the outer coma is dominated by smaller particles which contribute primarily to the coma-integrated ground-based observations. This is in agreement with \citet{Bertini2018} as they show that the backscattering ratio, which is proportional to the size of the dust particles, decreases with the distance $< 400$\,km from the nucleus.  Further, the ground-based polarimetric images of the coma \citep{Rosenbush2017} imply that the coma is inhomogeneous, i.e., different types of particles populate different regions in the coma. Polarization drops suddenly as the distance from the nucleus increases and then it starts to rise again reaching the near nucleus value at the outermost region of the coma and the tail. Be that as it may, inhomogeneities in the coma are not well understood and the explanation would also require sophisticated dynamical modelling of dust particles in the coma but this is out of the scope of the present work. 

Finally, the nucleus phase function is consistent with the coma phase function, i.e., particles in the near coma explain the nucleus phase function when packed densely on a rough surface. This means that multiple scattering has a significant contribution to the nucleus phase function. We have modelled the surface roughness as the statistical GRS shape, and linking this roughness to the real surface roughness of the nucleus is not straightforward. Using the real shape model of the nucleus would still require including an additional micro-roughness parameter to the model.

%{\bf JA: You could also discuss (either here or in Sec. 4, the expected difference or similarity between the GRS model of the nucleus and using the actual shape model covered in dust.}

\section{Conclusions} 

We have introduced a particle model that fits the coma and the nucleus phase functions at the visible wavelengths measured by the OSIRIS camera. The particles are modelled as aggregates composed of submicrometer-sized organic grains and micrometer-sized silicate grains. The model suggests that the dominant particle size in the coma varies between different measurements from 5\,\textmu m to 100\,\textmu m. The nucleus phase function can be explained with the same particles as in the coma packed densely on a rough surface. %To obtain more detailed constraints on the porosity and morphology of the grains, requires accurate measurements and modelling of the scattering features near the backscattering direction.  

\acknowledgments{{\em{Acknowledgments:}} This work has been funded by the ERC Starting Grant No. 757390 and the ERC Advanced Grant No. 320773. Computational resources have been provided by Gesellschaft f\"{u}r wissenschaftliche Datenverarbeitung mbH G\"{o}ttingen. The authors would like to thank N. Masoumzadeh and I. Bertini for sharing the phase function data. }


%\vspace{5mm}
%\facilities{HST(STIS), Swift(XRT and UVOT), AAVSO, CTIO:1.3m,


\software{FaSTMM \citep{Markkanen2017},  
          R$^2$T$^2$ \citep{Muinonen2018,Markkanen2018},
          SIRIS4 \citep{Lindqvist2018, Martikainen2018}}

\end{document}